\documentclass[10,conference,compsocconf]{IEEEtran}

\usepackage[utf8]{inputenc}
\usepackage[T1]{fontenc}
\usepackage[english]{babel}

\usepackage{cite}
\usepackage{url}
\usepackage[pdftex]{graphicx}
\usepackage[font=footnotesize]{caption}
\usepackage[font=footnotesize]{subcaption}

\begin{document}

\author{\IEEEauthorblockN{Vincent Primault, Sonia Ben Mokhtar, Lionel Brunie}
\IEEEauthorblockA{Université de Lyon, CNRS\\
INSA-Lyon, LIRIS, UMR5205, F-69621, France\\
\{vincent.primault,sonia.ben-mokhtar,lionel.brunie\}@liris.cnrs.fr}}

\title{Privacy-preserving Publication of Mobility Data with High Utility}

\maketitle

\begin{abstract}
An increasing amount of mobility data is being collected every day by different means, e.g., by mobile phone operators.
This data is sometimes published after the application of simple anonymization techniques, which might lead to severe privacy threats.
We propose in this paper a new solution whose novelty is two-fold.
Firstly, we introduce an algorithm designed to hide places where a user stops during her journey (namely points of interest), by enforcing a constant speed along her trajectory.
Secondly, we leverage places where users meet to take a chance to swap their trajectories and therefore confuse an attacker.
\end{abstract}


\section{Introduction}

The widespread adoption of location-aware devices such as smartphones has dramatically increased the quantity of mobility data that is being continuously collected.
However, collecting and sharing mobility data raises serious privacy concerns.
Among the known threats is the extraction of users'\textit{points of interest} (POIs)~\cite{Gambs11}, which can be defined as places where individuals regularly spend some time, e.g., home, work, a cinema or a mall.
By studying the semantics of these places, it is possible to infer sensitive knowledge like religious or political preferences.
Learning users' POIs can ultimately lead to learn about the real identity of individuals with a good accuracy.
Nevertheless, mobility data is still very valuable.
Publishing such information allows researchers to perform real-time traffic predictions, find out interesting patterns or discover social tendencies.
It is still an open and challenging issue to publish mobility traces of a set of users in a privacy-preserving manner.
To reach this objective, a classical solution that is applied in the literature is to alter user's geographical locations (e.g.,~\cite{Andres13,Abul10}).
However, this also alters the utility of published data, as trajectories are heavily distorted.
This is why we propose a new solution for privacy-preserving mobility data publishing that hides users' POIs.
Our challenge is to minimize the distortion of the geographical information contained in the published mobility traces.
To reach this objective, our solution distorts time and opportunistically swap trajectories of users when it is possible.

The remaining of this paper is structured as follows. We present related work in Section~\ref{sec:related}, an overview of our solution in Section~\ref{sec:solution}, before concluding in Section~\ref{sec:conclusion}.

\section{Related work}
\label{sec:related}

In 2002, Sweeney introduced the concept of $k$-anonymity.
The main idea behind it is that, for each quasi-identifier of a published dataset, there must be at least $k$ persons with this same quasi-identifier.
Abul et al. proposed \textit{Wait For Me}~\cite{Abul10}, which enforces $(k,\delta)$-anonymity for spatial data.
Their goal is to guarantee that at every instant there is at least $k$ users at a maximum distance $\delta$ of the others.
Their solution was shown to perform well with a synthetic dataset but having more difficulties to maintain a correct utility with a real-life dataset.

Differential privacy is a more recent concept introduced by Dwork in 2006.
The basic idea is that an aggregate result over a database should be almost the same whether or not a single row is present inside the database.
Andres et al. extended differential privacy through the concept of \textit{geo-indistinguishability}~\cite{Andres13}, which is designed to protect location data.
Like differential privacy, its strength is to provide formal and provable privacy guarantees.
However, it is not yet suitable to protect mobility datasets.
We have shown in~\cite{Mapomme14} that, on a real-life dataset, it does not prevent the extraction of at least 60~\% of the POIs even with a high privacy level.

Hoh et al.~\cite{Hoh05} proposed a solution to defeat multi-target tracking attacks.
These attacks are used against mobility data protected by removing the user identifier to link back together data that is likely to belong to a same user.
The proposed solution is to force paths of users to cross in areas where users are close, thus introducing confusion for an attacker.
The solution has only been tested against randomized movement models.

\section{Overview of our solution}
\label{sec:solution}

\begin{figure*}
        \centering
        \begin{subfigure}[b]{0.3\textwidth}
                \includegraphics[width=5cm,height=4cm]{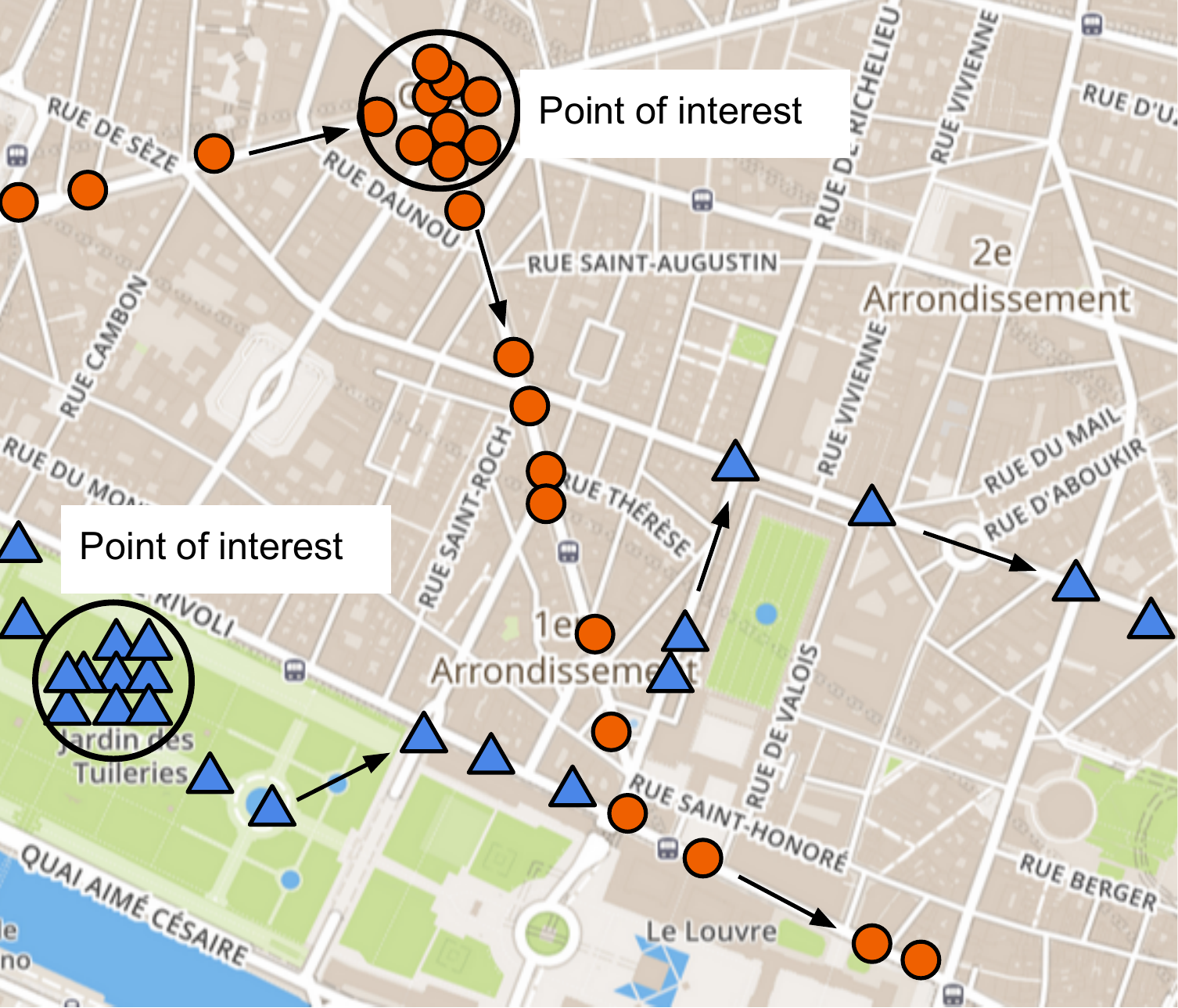}
                \caption{Original trace}
                \label{fig:original}
        \end{subfigure}%
        ~ 
        \begin{subfigure}[b]{0.3\textwidth}
                \includegraphics[width=5cm,height=4cm]{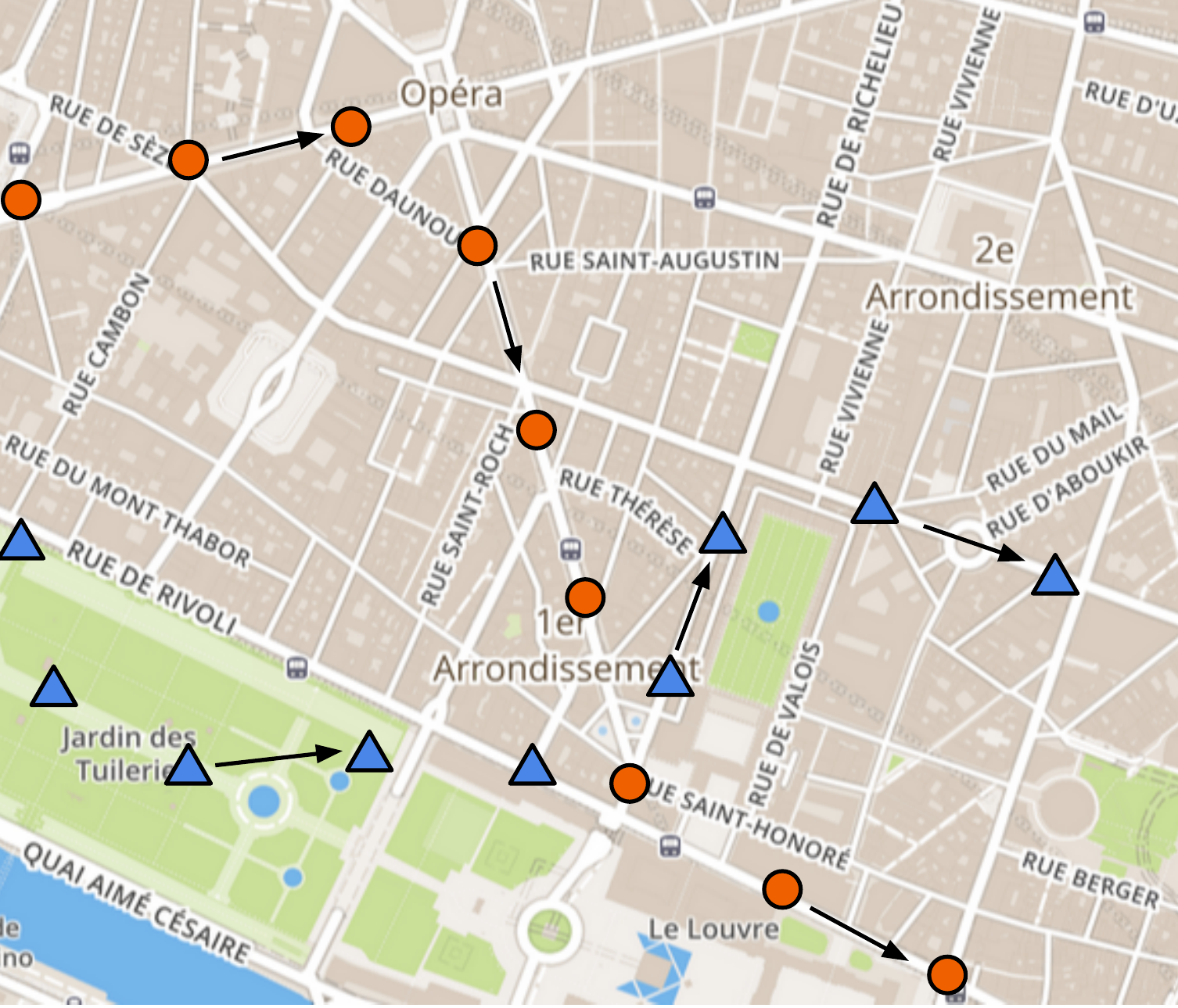}
                \caption{After enforcing a constant speed}
                \label{fig:smoothing}
        \end{subfigure}
        ~ 
        \begin{subfigure}[b]{0.3\textwidth}
                \includegraphics[width=5cm,height=4cm]{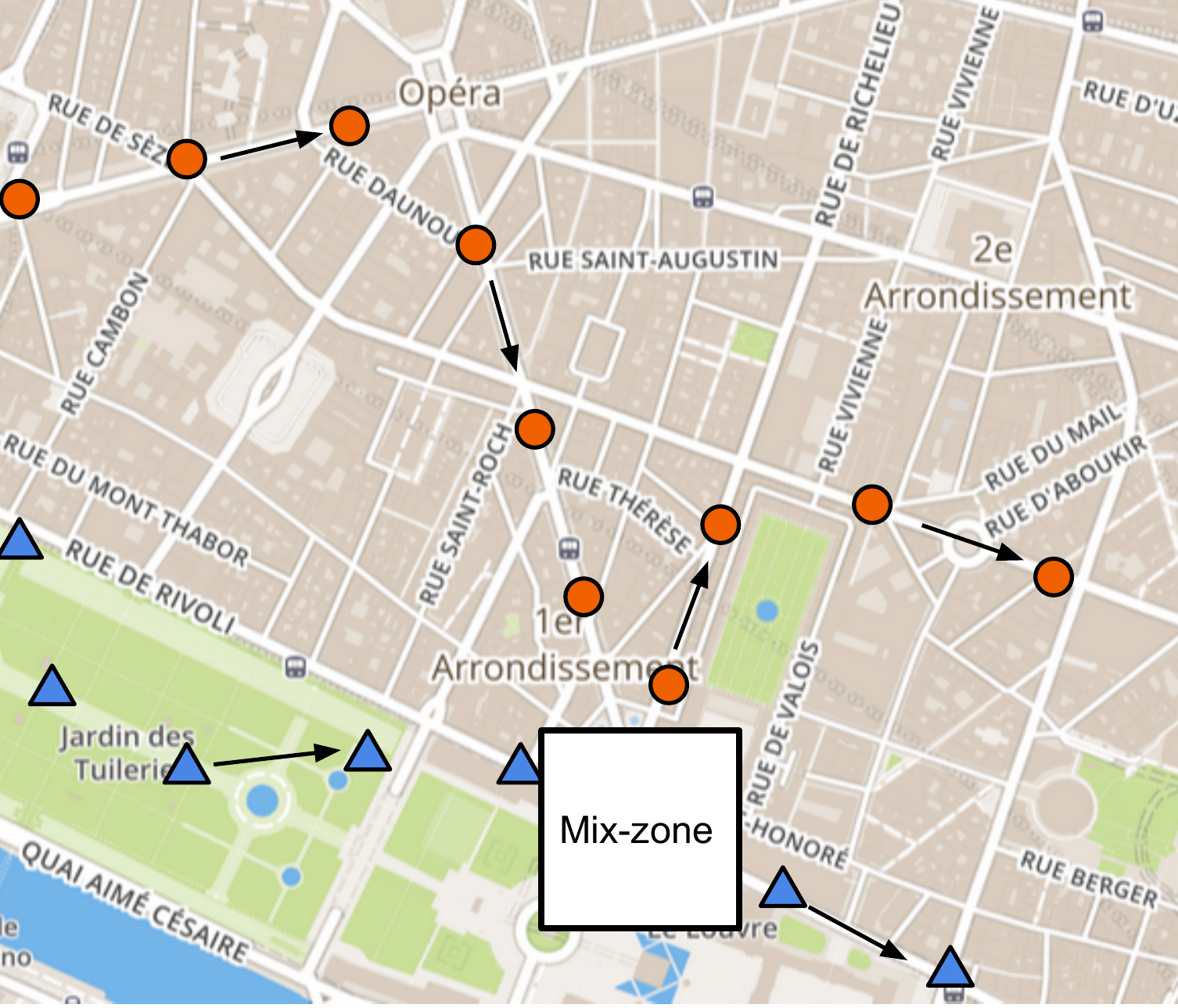}
                \caption{After swapping trajectories}
                \label{fig:confusion}
        \end{subfigure}
        \caption{An example of two mobility traces anonymized with our solution.}\label{fig:overview}
\end{figure*}

The first privacy threat we want to address is the extraction of users' POIs.
These correspond to places where users stop and spend some time, before moving to another place.
A trace is basically a list of POIs, that can be viewed as clusters of points (as shown on Fig.~\ref{fig:original}), with transitions in between.
To hide users' POIs, we propose to enforce a constant speed along the mobility trace of each user.
More specifically, we transform mobility traces to enforce an equal duration and distance between two consecutive points.
If we can guarantee that speed is constant throughout a trace, it becomes difficult for an adversary to spot where a user stopped because there is no point at which she appears to be stationary.
Clues can still be obtained from background knowledge (e.g. the probability is higher to stop in a park that in the middle of a motorway) but there will be no certainty for an attacker (e.g. a user can either have just crossed the park or had a picnic there).
Figure~\ref{fig:smoothing} shows the result of this operation applied on two mobility traces.
From this figure, we can see that the POIs of the users have been hidden and that points on each trace are evenly distributed.

The other privacy threat we want to address in this paper is the re-identification of users.
To reach this objective, we exploit natural paths crossings to confuse an attacker. 
When users move during a day, they continuously meet other users in public transportations, malls, work places, etc.
These meeting areas are called mix-zones, as introduced in~\cite{Beresford03}.
Mix-zones are well-delimited areas where nobody is tracked; we only know where and when users enter and leave a mix-zone, without any insight about what happens inside.
Before users leave a mix-zone, their identifiers are possibly shuffled.
A user entering a mix-zone labelled as "user A" could either leave it labelled as "user B" or remain "user A".
It therefore helps breaking the correlation between traces before and after the mix-zone.
We do not want to artificially distort traces to force users to meet, but instead we take advantage of existing mix-zones.
Figure~\ref{fig:confusion} shows that the two traces have been swapped inside the mix-zone.

Our main utility goal was to minimally distort the location.
The first step introduces only error when interpolating new points between known ones.
If the sampling rate is high enough, this interpolation should be precise enough to introduce almost no spatial inaccuracy.
Similarly, the second step only swap user identifiers but does not alter the location.
The only utility loss comes from the fact we suppress points inside mix-zones, but this should be a reasonable degradation as long as mix-zones remain reasonably small.
Tough, we acknowledge not all queries can be implemented with our solution.
For example, studying transitions between locations cannot be done because users can possibly be swapped.
We believe there is not one solution that fits all use cases, and that the appropriate solution must be chosen according to both users' preferences in terms of privacy and the analysts' objectives.

\section{Conclusion}
\label{sec:conclusion}

In this paper we presented an overview of a new solution to anonymize mobility datasets.
Its novelty resides in the fact that it obfuscates time instead of obfuscating location, which allows to have a better spatial accuracy than state of the art solutions.
Furthermore, our solution swaps trajectories in order to confuse the attacker, without compromising utility.
We now plan to evaluate our solution with real-life datasets and study its resiliency against common privacy attacks and the practical utility it can preserve.

\section*{Acknowledgment}
This work was supported by the LABEX IMU (ANR-10-LABX-0088) of Université de Lyon, within the program "Investissements d'Avenir" (ANR-11-IDEX-0007) operated by the French National Research Agency (ANR).

\bibliographystyle{IEEEtran}
\bibliography{IEEEabrv,bibli}

\begin{thebibliography}{1}
\providecommand{\url}[1]{#1}
\csname url@samestyle\endcsname
\providecommand{\newblock}{\relax}
\providecommand{\bibinfo}[2]{#2}
\providecommand{\BIBentrySTDinterwordspacing}{\spaceskip=0pt\relax}
\providecommand{\BIBentryALTinterwordstretchfactor}{4}
\providecommand{\BIBentryALTinterwordspacing}{\spaceskip=\fontdimen2\font plus
\BIBentryALTinterwordstretchfactor\fontdimen3\font minus
  \fontdimen4\font\relax}
\providecommand{\BIBforeignlanguage}[2]{{%
\expandafter\ifx\csname l@#1\endcsname\relax
\typeout{** WARNING: IEEEtran.bst: No hyphenation pattern has been}%
\typeout{** loaded for the language `#1'. Using the pattern for}%
\typeout{** the default language instead.}%
\else
\language=\csname l@#1\endcsname
\fi
#2}}
\providecommand{\BIBdecl}{\relax}
\BIBdecl

\bibitem{Gambs11}
S.~Gambs, M.-O. Killijian, and M.~N. del Prado~Cortez, ``{Show Me How You Move
  and I Will Tell You Who You Are},'' \emph{Transactions on Data Privacy},
  vol.~4, no.~2, pp. 103--126, Aug. 2011.

\bibitem{Andres13}
M.~E. Andr{\'e}s, N.~E. Bordenabe, K.~Chatzikokolakis, and C.~Palamidessi,
  ``{Geo-indistinguishability: Differential Privacy for Location-based
  Systems},'' in \emph{{Proceedings of CCS'13}}.\hskip 1em plus 0.5em minus
  0.4em\relax ACM, 2013, pp. 901--914.

\bibitem{Abul10}
O.~Abul, F.~Bonchi, and M.~Nanni, ``Anonymization of moving objects databases
  by clustering and perturbation,'' \emph{Information Systems}, vol.~35, no.~8,
  pp. 884--910, 2010.

\bibitem{Mapomme14}
V.~Primault, S.~Ben~Mokhtar, C.~Lauradoux, and L.~Brunie, ``{Differentially
  Private Location Privacy in Practice},'' in \emph{{Proceedings of MOST'14}},
  May 2014.

\bibitem{Hoh05}
B.~Hoh and M.~Gruteser, ``Protecting location privacy through path confusion,''
  in \emph{Proceedings of SECURECOMM'05}.\hskip 1em plus 0.5em minus
  0.4em\relax IEEE Computer Society, 2005, pp. 194--205.

\bibitem{Beresford03}
A.~Beresford and F.~Stajano, ``Location privacy in pervasive computing,''
  \emph{Pervasive Computing, IEEE}, vol.~2, no.~1, pp. 46--55, Jan. 2003.

\end{thebibliography}

\end{document}